\documentclass[twocolumn]{aastex701}
\usepackage{CJK}

\usepackage{graphicx}	
\usepackage{amsmath}	
\def\specSN{6800} 

\def\womasssystpercent{42.9}
\def\wamasssystpercent{49.6}

\def\ninety{19.27}
\def\fifty{20.61}

\def\lmninety{19.35}
\def\lmfifty{20.21}

\def\hmninety{19.25}
\def\hmfifty{20.90}

\def\masspercent{22\% and 327\%}

\def\blueninety{19.57}
\def\bluefifty{20.84}

\def\redninety{19.55}
\def\redfifty{20.18}

\def\colorpercent{24.2\% and 975\%}

\def\groundredninety{18.98}
\def\groundredfifty{23.09}
\def\groundblueninety{19.51}
\def\groundbluefifty{23.26}

\def\zgrismonly{1.05}
\def\zgrismnum{3500}
\def\zgroundonly{0.84}
\def\zgroundnum{2900}
\def\zprismonly{1.20}
\def\zprismnum{2900}

\def\womasssyst{-0.0223$\pm$0.0065}

\def\wamasssyst{0.1062$\pm$0.0318}

\def\masssyst{0.0017$\pm$0.0005}
\def\masssystRSD{0.0025}

\def\wocolorsyst{0.0066$\pm$0.002}

\def\wacolorsyst{-0.0266$\pm$0.0079}

\def\colorsyst{0.0005$\pm$0.0002}
\def\colorsystRSD{0.0009}

\def\woshiftsyst{-0.0032$\pm$0.0009}

\def\washiftsyst{0.0142$\pm$0.004}

\def\shiftsyst{0.0000$\pm$0.0000}
\def\shiftsystRSD{0.0000}

\def\wstatuncert{0.018}
\def\wostatuncert{0.052}
\def\wastatuncert{0.214}

\begin{document}


\title[\textit{Roman} grism for SNIa]{Characterizing the \textit{Roman} Grism Redshift Efficiency of Type Ia Supernova Host Galaxies for the High-Latitude Time-Domain Survey}

\author[orcid=0000-0003-3917-0966]{Rebecca~C.~Chen}
\affiliation{Department of Physics, Duke University, Durham, NC 27708, USA}
\email[show]{rcchen0@stanford.edu}  

\begin{CJK*}{UTF8}{gbsn}
\author[0000-0001-9557-9171]{Zhiyuan Guo(郭致远)}
\affiliation{Department of Physics, Duke University, Durham, NC 27708, USA}
\email{}

\author[0000-0002-4934-5849]{Dan Scolnic}
\affiliation{Department of Physics, Duke University, Durham, NC 27708, USA}
\email{} 

\author{Bhavin Joshi}
\affiliation{Department of Physics and Astronomy, Johns Hopkins University, Baltimore, MD 21218, USA}
\email{} 

\author{Richard Kessler}
\affiliation{Kavli Institute for Cosmological Physics, University of Chicago, Chicago, IL 60637, USA}
\affiliation{Department of Astronomy and Astrophysics, University of Chicago, Chicago, IL 60637, USA}
\email{} 

\author[orcid=0000-0002-1296-6887]{Llu\'is Galbany}
\affiliation{Institute of Space Sciences (ICE-CSIC), Campus UAB, Carrer de Can Magrans, s/n, E-08193 Barcelona, Spain}
\affiliation{Institut d'Estudis Espacials de Catalunya (IEEC), 08860 Castelldefels (Barcelona), Spain}
\email{l.g@csic.es} 

\author{Rebekah Hounsell}
\affiliation{University of Maryland, Baltimore County, 1000 Hilltop Cir, Baltimore, MD 21250}
\affiliation{NASA Goddard Space Flight Center, 8800 Greenbelt Rd, Greenbelt, MD 20771}
\email{}

\author[0000-0002-2313-5763]{Diane~M.~Markoff}
\affiliation{Department of Mathematics and Physics, NC Central University, Durham, NC 27707, USA}
\email{} 

\author[orcid=0000-0002-1873-8973]{Benjamin~M.~Rose}
\affiliation{Department of Physics and Astronomy, Baylor University, One Bear Place \#97316, Waco, TX 76798-7316, USA}
\email{Ben_Rose@baylor.edu} 

\author[orcid=0000-0001-5402-4647]{David Rubin}
\affiliation{Department of Physics and Astronomy, University of Hawai'i at Manoa, Honolulu, HI 96822}
\affiliation{Lawrence Berkeley National Laboratory, 1 Cyclotron Road, Berkeley, CA 94720, USA}
\email{} 

\collaboration{all}{The Roman Supernova Cosmology Project Infrastructure Team}

\begin{abstract}

The High-Latitude Time-Domain Survey (HLTDS) for the \textit{Nancy Grace Roman Space Telescope} (\textit{Roman}) will discover thousands of high redshift Type Ia supernovae (SNeIa) to make generation-defining cosmological constraints on dark energy. To construct the \textit{Roman} SN Hubble diagram, a strategy to obtain redshifts must be determined. While the nominal HLTDS will use only the \textit{Roman} prism, in this work we consider the utility of the \textit{Roman} grism observations from overlap with the High-Latitude Wide-Area Survey for SNIa cosmology. We determine a galaxy grism redshift recovery rate by simulating dispersed grism images and measuring redshifts with the \texttt{Grizli} software, obtaining an $H$-band 50\% redshift recovery at magnitude \fifty\ and 90\% recovery at magnitude \ninety. To estimate the total number of spectroscopic redshifts expected for \textit{Roman} SN cosmology, we also consider a \textit{Roman} prism SN redshift efficiency and a ground-based telescope redshift efficiency for host-galaxies. We apply these redshift efficiencies to SNIa catalog level simulations and predict that $\sim$\specSN\ SNe will have a SN or host spectroscopic redshift. Second, we evaluate the size of potential systematics related to modeling the grism redshift efficiency by considering the impact of additional dependencies on stellar mass and host galaxy color. We estimate the largest potential size of this systematic to be \wocolorsyst{} and \wacolorsyst{}, roughly \womasssystpercent\ and \wamasssystpercent\% of the statistical uncertainty for $w_0$ and $w_a$ respectively. Lastly, we consider the effects of assuming different redshift sources on the HLTDS survey strategy optimization by measuring relative changes to the dark energy Figure of Merit. 

\end{abstract}



\section{Introduction}\end{CJK*}

The \textit{Nancy Grace Roman Space Telescope} (henceforth \textit{Roman}; \citealt{Spergel15}) is NASA's next flagship space mission, poised to launch by October 2026. One key scientific aim of $Roman$ is to answer long-standing questions in cosmology, particularly regarding the nature and evolution of dark energy. To do so, $Roman$ will use a primary mirror 2.4 m in diameter and a Wide Field Instrument (WFI) consisting of 9 near-infrared (NIR) imaging filters, a prism, and a grism. More specifically, the two slitless-spectroscopy dispersive elements are a low-resolution prism (P127; $R\approx80-180$; 0.75-1.80 microns) and a higher-resolution grism (G150; $R\approx460$; 1.00-1.93 microns). With these facilities, the mission will carry out a High-Latitude Wide-Area Survey (HLWAS) with an imaging component for weak lensing and large-scale structure studies and a spectroscopic component to measure baryon acoustic oscillations (BAO) and redshift-space distortions (RSD), as well as a High-Latitude Time-Domain Survey (HLTDS) for transient studies, including Type Ia supernova (SNIa) cosmology. The field of view (FoV) is 0.281 square degrees, 100 times larger than the Hubble Space Telescope, which will allow for an unprecedented number of SNIa discovered, particularly at redshift $z > 1$ \citep{Hounsell18, Akeson19}. 

To make precise constraints of the dark energy equation-of-state parameter $w$ or other parameterizations such as $w_0$-$w_a$ using SNeIa, redshifts of the supernova or preferably, host galaxy, are required. Major SNIa analyses to date \citep{JLA, Scolnic18, PantheonPlus, Vincenzi24} have relied exclusively on spectroscopic follow-up programs to obtain as many redshifts as possible for a given SN sample. Realistically, well-measured redshifts will be obtainable for only a subset of the order $10^4$ SNe that will be observed by $Roman$ \citep{Rose21}. This introduces a biased sub-selection of SN host galaxies, as brighter galaxies will be preferentially observed. This type of redshift selection effect must be well modeled and accounted for to estimate bias corrections with simulations. The recent Dark Energy Survey 5-year SN analysis (DES-SN5YR) modeled this efficiency ($\epsilon_{z_{\rm spec}}$) of obtaining a host spectroscopic redshift (spec-$z$) as a four-dimensional function of field, host galaxy brightness, color, and year of SN discovery \citep{Vincenzi21a, Vincenzi24}. While efforts to develop fully photometric analysis methods (without requiring spectroscopic redshifts) are underway \citep{Chen22, RuhlmannKleider22, Mitra23, Chen25}, a complete cosmological analysis with systematics has not yet been performed with data. An understanding of the spectroscopic redshift efficiency will still be a crucial simulation component both for a ``standard'' spec-$z$ analysis and for analyses that use both spec-$z$ and photometric redshifts (photo-$z$).

Reference surveys have been proposed for the HLWAS \citep{Wang22} and HLTDS \citep{Hounsell18, Rose21} and recommendations have been made from the $Roman$ Observation Time Allocation Committee (ROTAC)\footnote{\url{ https://asd.gsfc.nasa.gov/roman/comm_forum/forum_17/Core_Community_Survey_Reports-rev03-compressed.pdf}} for each Core Community Survey. The HLTDS Definition Committee builds on the Design Reference Survey from \citet{Rose21} and recommends that $\sim 20\%$ of the HLTDS time be dedicated to prism spectroscopy. The choice to use the prism over the grism for the HLTDS is due to the prism's higher throughput ($\sim$ 3 mag). This is more desirable for observing fainter SNe with broad spectral features which do not require high resolution. Efforts to understand the efficiency of recovering SN redshifts have been undertaken as part of the previous $Roman$ Science Investigation Team SN cosmology efforts. \cite{Joshi22} generated simulated prism observations and measured the SN redshift recovery as a function of exposure time. They found that with a requirement of $\sigma_z = (|z - z_{\rm true}|)/(1+z)\leq 0.01$, the longest exposure times of 3 hours can reach 50\% efficiency as faint as $\sim25.5$. \cite{Rubin22} emphasizes the importance of the prism for the HLTDS to improve subclassification of supernovae and to constrain population distributions as a function of redshift.

Notably, the HLTDS reference survey also makes the simplifying assumption that precise redshifts will be available for the entire $Roman$ SN sample. While some SN and host galaxy redshifts will be measured with the prism, the rest are assumed to come from partner spectroscopic follow-up programs on ground-based telescopes such as the Subaru Telescope Prime Focus Spectrograph (PFS; \citealt{PFS}), located in Hawaii. An avenue that has not yet been fully explored is the utility of the $Roman$ grism, which will be primarily used by the HLWAS and will overlap with the Southern component of the HLTDS (EDFS; Euclid Deep Field South). Characterizing the redshift efficiency of the grism for SN host galaxies will also allow for more informed decisions regarding priority in external spectroscopic follow-up resources.

Several works have already simulated 2D dispersed images for the $Roman$ grism, including for the spectroscopic HLWAS reference survey \citep{Wang22} and specifically for red, quiescent galaxies \citep{Guo25}. \cite{Wang22} used the aXeSIM software \citep{aXeSIM} developed for the Hubble Space Telescope to simulate images and spectra to forecast constraints and systematics for the HLWAS. \cite{Guo25} instead use the Grism and Redshift Line Analysis software (\verb|Grizli|; \citealp{Grizli}) on direct images generated by the OpenUniverse2024 effort \citep{Troxel23, OpenUniverse24}. \citet{Gabrielpillai24} also present ESpRESSO, another \textit{Roman} grism simulation framework that delivers grism simulations with a more complete suite of roll angles and dithers for deep grism observations.

In this work, we focus on the usefulness of the $Roman$ grism, rather than the prism, for supernova cosmology, particularly for obtaining redshifts of SN host galaxies. To estimate the grism redshift recovery rate, we build on the results of \cite{Guo25} to simulate grism spectroscopy following the reference HLWAS and analyze the 2D spectra using the \texttt{Grizli} software. We further consider approximate redshift efficiencies for the \textit{Roman} prism and ground-based spectroscopic follow-up programs. Using these redshift efficiencies as inputs to our catalog-level SN simulations, we simulate the reference HLTDS survey and provide initial estimates for the number of spectroscopic redshifts that will be available for the \textit{Roman} SNIa cosmological sample. We then use these HLTDS simulations to evaluate the size of potential systematics associated with modeling the grism redshift efficiency and consider the impact of different redshift source assumptions on survey design and optimization.

The paper is organized as follows. In Section \ref{sec:efficiencies} we describe the grism simulations and present the overall efficiency of recovering host-galaxy redshifts using the $Roman$ grism. In Section \ref{sec:sn} we describe the catalog simulations of SNeIa and their host galaxies, describe approximate prism and ground-based efficiencies, and present estimates for a $Roman$ SN cosmology sample. In Section \ref{sec:systematics} we discuss several options for a redshift efficiency modeling systematic. In Section \ref{sec:FOM} we discuss the implications of assuming different redshift sources on the HLTDS survey optimization and dark energy figure-of-merit for various survey configurations. Lastly, we summarize and conclude in Section \ref{sec:conclusion}.

\section{Grism Simulations and Redshift Efficiency}\label{sec:efficiencies}
 Here we provide a brief overview of the grism simulations used to measure a general galaxy redshift recovery rate following \citet{Guo25} (henceforth G25). While G25 focuses on measuring the grism efficiency for only red, quiescent galaxies, here we extend the general methodology to include all types of galaxies. We direct readers to \citet{Guo25} for more comprehensive details on the method and validation.

 \subsection{Grism simulations}\label{sec:grismsims}

\begin{figure*}
    \includegraphics[width=\linewidth]{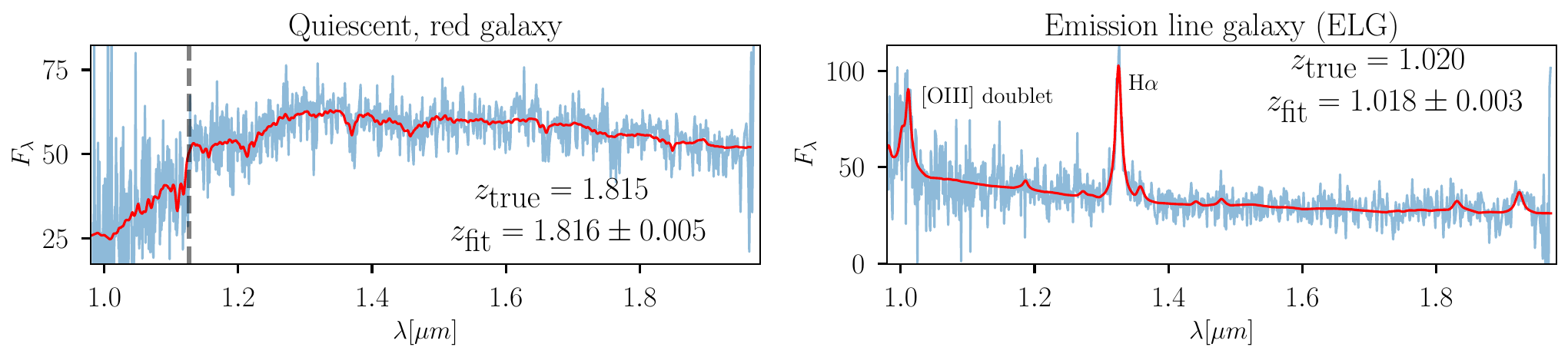}
    \caption{Spectra extraction and template fitting examples for a quiescent, red galaxy (left panel) and an emission line galaxy (right panel). The blue solid lines show the optimally extracted spectra and the red solid line shows the \texttt{Grizli} best-fit template. The true redshift and measured redshift are shown in each panel. The 4000\AA\ break is marked with dashed grey in the left panel, and strong emission lines are labeled in the right panel. The flux unit is in $10^{-19}$ erg/s/cm$^2$/\AA.}
    \label{fig:extraction_ex}
\end{figure*}

We utilize the \texttt{Grizli} software to simulate $Roman$ grism images, generating grism outputs from direct images, referred to as reference images, alongside corresponding segmentation maps, which designate individual sources. For each source, the flux within its segmentation region is dispersed according to the grism's dispersive direction, with the segmentation size dictating the width of the simulated spectrum. This study employs simulated direct images in the F158 band from the OpenUniverse2024 simulation \citep{OpenUniverse24}, which spans an overlapping region between the Rubin Observatory Legacy Survey of Space and Time (LSST) Wide-Fast-Deep survey and $Roman$ footprints. To simulate grism images, we require a spectral energy distribution (SED) for each source. OpenUniverse2024 utilizes the extragalactic model \texttt{Diffsky} \citep{Diffstar, DSPS}\footnote{\url{https://lsstdesc-diffsky.readthedocs.io/en/latest/}} to generate the input SEDs for all galaxies in the direct image simulation, and the same corresponding SEDs are assigned to each source in the grism simulation.

Our grism simulation parameters follow the guidelines set by the spectroscopic HLWAS reference survey \citep{Wang22}. We produce grism images at four position angles (0, 5, 170, and 175 degrees) with two exposures at each angle, each lasting 347 seconds. It is important to note that we do not replicate the exact dithering and tilting pattern proposed in \citet{Wang22}, as the primary purpose of our simulations is to measure the redshift recovery rate based on expected exposure times. All roll angles are centered on the same field, implying that all sources are positioned identically on the detector for all eight grism observations, accumulating a total exposure time of 2776 seconds per source. The impact of this assumption is explored in \citet{Guo25}, which concludes that each lost exposure causes a decrement of approximately 0.1 mag in the 50\% efficiency level. 

The original G25 simulations incorporated the noise model of \citet{Joshi22}, including a zodiacal background of 1.047 electrons/pixel/second, a dark current of 0.0015 electrons/pixel/second, and a readout noise RMS of 8 electrons. We updated the noise model to be consistent with more recent measurements and image processing parameters. The zodiacal and thermal background is estimated (conservatively) at 0.937 electrons/pixel/second based on a tophat approximation to the grism passband when including all orders. The read noise is typically 5-6 DN\footnote{\url{https://roman-docs.stsci.edu/roman-instruments-home/wfi-imaging-mode-user-guide/wfi-detectors/detector-performance/instrumental-noise}}, corresponding to 9-10 electrons. We also adjusted the noise estimate to account for the benefit of the slope-fitting process, which reduces the effective read noise by combining multiple reads. For the relevant exposure times and signal levels, the signal-to-noise ratio can be estimated to within 5\% by adopting an effective read noise 0.63 times the single-read noise. The effective read noise is thus taken to be 6e- and is added in quadrature with the Poisson noise contribution.

We acknowledge the omission of a wavelength-dependent point spread function (PSF) in our grism simulations and restrict our simulation to only the first spectral order, or the science order (1-1). As discussed in G25, contamination from source light in other orders is expected to be minimal in most scenarios. However, diffuse sky light in other orders does contribute to the total background. This contribution is therefore included in our updated zodiacal contribution. The investigation of the effects of a wavelength-dependent PSF and contamination from additional spectral orders is left for future research.

\subsubsection{Redshift extraction}
The \texttt{Grizli} framework uses a two-step redshift determination process. Initially, redshifts are coarsely estimated using a predefined grid and a set of continuum and emission line templates. This is followed by a refined fitting procedure focused around the chi-squared minimum, utilizing a tighter redshift grid to enhance accuracy.  The redshift fitting is conducted directly on contamination-subtracted 2D cutout frames for each source, effectively addressing spectral smearing and self-contamination. For each fitted spectrum, the final result includes the redshift probability distribution function on the refined redshift grid, $P(z)$, the best-fit redshift, corresponding uncertainties, and the strength of emission lines if any are detected.

Moreover, \texttt{Grizli} facilitates the extraction of 1D spectra from these 2D frames, incorporating data from multiple exposures across all position angles. This integration employs the optimal extraction algorithm from~\citet{Horne86}. The resulting 1D spectra are used both for visual evaluation of the fit quality and for calculating the signal-to-noise ratio (S/N) using the DER-SNR algorithm~\citep{Stoehr08}. Additionally, \texttt{Grizli}'s redshift fitting results have been validated by comparison with those from other independent spectroscopic redshift estimation packages, such as \texttt{Bagpipes}\footnote{\url{https://bagpipes.readthedocs.io/en/latest/}}. Preliminary comparisons, as detailed in G25, demonstrate strong agreement between the methods, reinforcing the reliability of \texttt{Grizli}'s redshift estimation approach. In Figure~\ref{fig:extraction_ex}, we show two examples of the extracted 1D spectra and corresponding redshift fitting results for a quiescent, red galaxy and an emission line galaxy randomly selected from the simulations. 

\subsection{Grism redshift efficiency}\label{sec:grismeff}
We consider a redshift measured successfully if one of the following sets of conditions are satisfied:

\begin{enumerate}
    \item the redshift residual $|(z_{\rm fit} - z_{\rm true})/(1+z_{\rm true})| < 0.02$,
    \item the S/N of the extracted spectra $\geq 5$, and
    \item the measured $p(z)$ has only one dominant peak.
\end{enumerate}
OR
\begin{enumerate}
    \item any two of the H$\alpha$, [OII], [OIII], SII, SIII, or H$\beta$ emission lines are detected to a typical emission-line flux limit of $10^{-16}$ erg $\textrm{s}^{-1} \textrm{cm}^{-2}$. 
\end{enumerate}

The first set of criteria is proposed in \citet{Guo25} to select well-measured and reliable redshifts for red, quiescent galaxies. The second selection criterion is applied to include the Emission Line Galaxies (ELGs) that are the primary target of the spectroscopic HLWAS, particularly at $1 < z < 3$. The emission line flux limit of $10^{-16}$ erg $\textrm{s}^{-1} \textrm{cm}^{-2}$ is the proposed $6.5\sigma$ depth for the $Roman$ spectroscopic HLWAS reference survey~\citep{Wang22}. To ensure reliable redshift estimates for ELGs, we require that at least two strong emission lines must be present in the extracted spectra. We define the redshift efficiency as the fraction of galaxies for which a redshift is measured successfully relative to all simulated galaxies. In Appendix A, we present the corresponding results for more relaxed conditions, such as requiring only one line detection and a lower detection threshold. This greatly improves the efficiency at the cost of estimating more uncertain redshifts.

\begin{figure}
    \centering
    \includegraphics[width=\columnwidth]{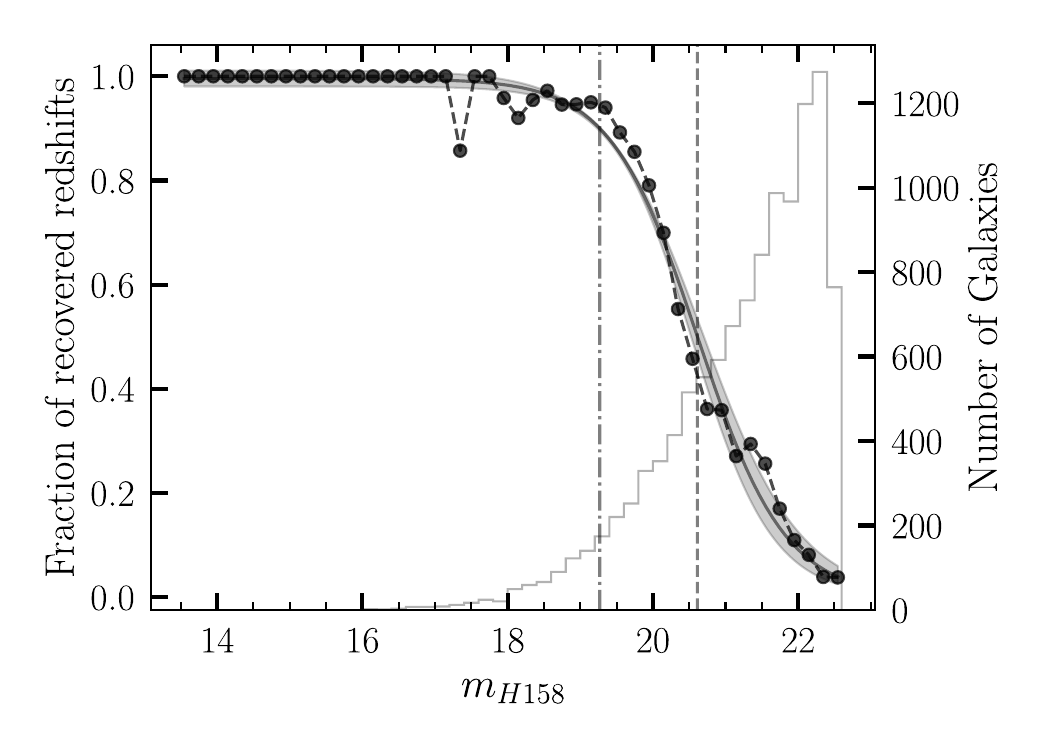}

    \includegraphics[width=\columnwidth]{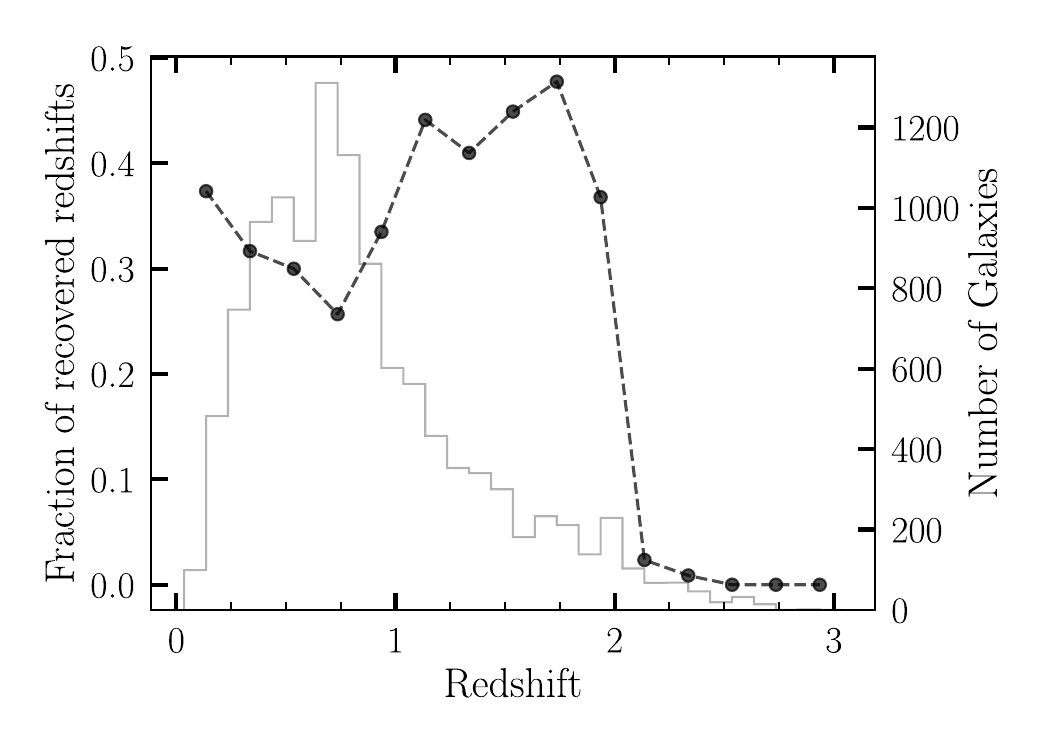}

    \includegraphics[width=\columnwidth]{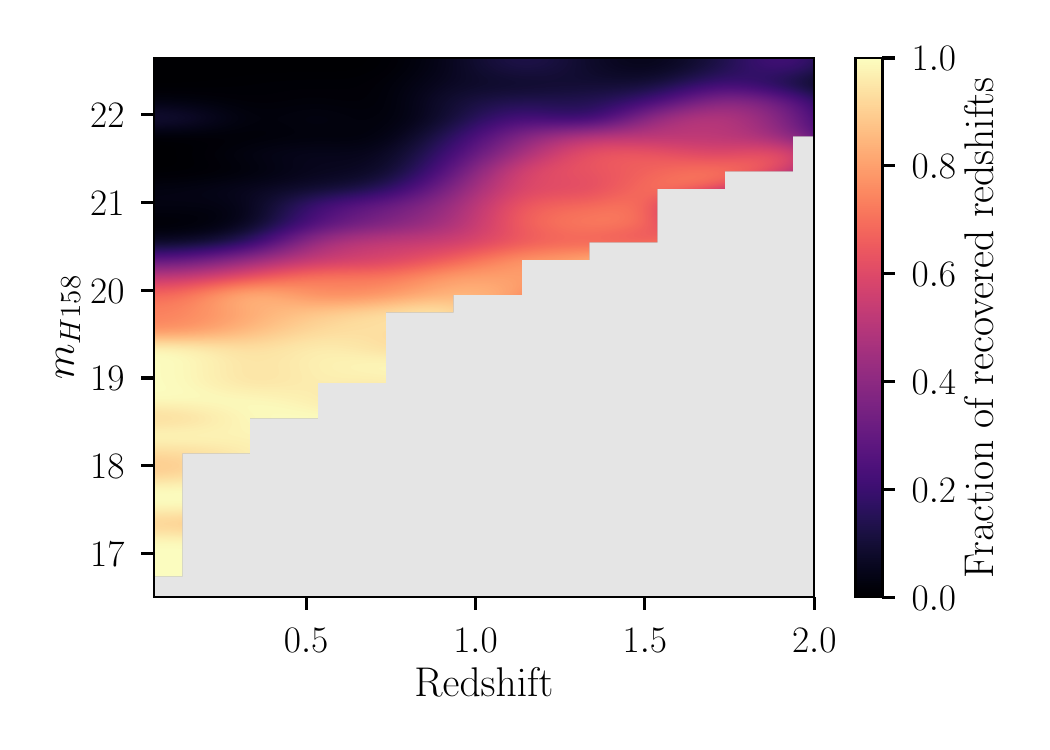}

    \caption{
    Top: Redshift efficiency from the grism image simulations as a function of $H$-band magnitude, with a logistic regression curve fitted to the data. The left y-axis indicates the fraction of redshift completeness (i.e., redshift efficiency). The background gray histogram shows the overall magnitude distribution of galaxies for which redshifts are successfully estimated, with the number of galaxies on the right y-axis. The 90\% and 50\% efficiencies are marked with dash-dotted and dashed lines, respectively.
    Middle: Redshift efficiency as a function of redshift, with a background gray histogram of galaxy redshifts. No fit is provided as the curve is illustrative and not used as an input to any SN simulations.
    Bottom: Heat map of redshift completeness in $H$-band magnitude and redshift space.
    }
    \label{fig:zeff}
\end{figure}

In the top panel of Figure \ref{fig:zeff}, we show the redshift efficiency as a function of F158 ($H$-band) magnitude measured from the grism simulations as presented in Section \ref{sec:grismsims}. We find 90$\%$ or greater completeness for magnitudes brighter than \ninety{} and 50$\%$ for magnitudes brighter than \fifty{}. In the middle panel of Figure \ref{fig:zeff}, we show the effective efficiency as a function of redshift. Notably, the efficiency does not peak until $z\sim1.3$. This is because the primary emission lines targeted by the spectroscopic HLWAS are H$\alpha$ and [OIII], which are covered by $Roman$'s grism bands only for $1.1 < z < 1.9$. For red galaxies, the 4000 \r{A} break is also covered by the grism filters at $z > 1.4$. We note also the steep drop-off in completeness at $z \sim 2$ which can be partially attributed to the wavelength range, as well as the fact that we choose not to simulate spectra for galaxies fainter than $m_\mathrm{F158} = 23$ to reduce computational cost. In the bottom panel of Figure \ref{fig:zeff}, we plot the redshift completeness as a heatmap in magnitude-redshift space.

\section{SN simulations}\label{sec:sn}

In this section, we apply the general grism redshift efficiency derived from \textit{image} simulations in the previous section to \textit{catalog level} SN simulations of the HLTDS. In Section \ref{sec:sim_sn_host}, we describe how we simulate SNe and their host galaxies. In Section \ref{sec:othereff}, we describe approximate efficiencies for the \textit{Roman} prism and ground-based telescopes as additional sources of host spectroscopic redshifts. Lastly, in Section \ref{sec:final_sn_effs}, we describe the overall redshift efficiency when combining the grism, prism, and ground-based spectroscopic resources to provide an approximate redshift schema for HLTDS SNIa cosmology and illustrate how further resources may be best allocated. 

\subsection{SN and host galaxy simulations}\label{sec:sim_sn_host}
To simulate SNIa and their host galaxies, we use the SuperNova ANAlysis Software (SNANA; \citealt{Kessler09}) orchestrated with the Pippin software \citep{Pippin}. Briefly, SNe are generated in three steps: i) A spectro-temporal model such as SALT3 \citep{Guy07, Kenworthy21} is used to produce a time-varying rest-frame SED, which is modified by various astrophysical effects such as dust reddening and redshifting. The SED is then integrated over filter passbands to obtain broadband fluxes. ii) Measurement noise specific to the survey observations is added (e.g. zeropoints). iii) Survey-dependent detection logic and detection efficiencies (e.g. our measured spectroscopic redshift efficiency) are applied. 

We use the SALT3 near-infrared extension (SALT3-NIR; \citealp{Pierel22}) to generate our SNIa light-curves. We model the SN intrinsic scatter model using the G10 \citep{Guy10} model and underlying parent populations for stretch and color from \cite{BBC}. We simulate only SNIa without any core collapse populations, as previous studies \citep{Vincenzi23, Vincenzi24} have indicated that with standard cosmological cuts and photometric classification the contribution of core collapse contamination to systematic uncertainties is minimal ($\sim3.5\%$).

Here we follow the reference survey design described in \cite{Rose21} as our baseline simulation. In brief, the survey is composed of two tiers, wide (in F062, F087, F106, F129; i.e. $R$, $Z$, $Y$, $J$) and deep (in F106, F129, F158, F184; i.e. $Y$, $J$, $H$, $K$), with $25\%$ spectroscopic time dedicated to the prism. The detection logic applied is to require Signal-to-Noise Ratio (SNR) of $>5$ for at least two filters.  

\subsubsection{Host galaxies}
To simulate the SN host galaxies, we use the host galaxy library (HOSTLIB) produced from the extra-galactic catalog covering the ELAIS-S1 Deep-Drilling Field used for the OpenUniverse2024 joint $Roman$-Rubin simulations \citep{OpenUniverse24}. The SEDs and associated photometry for these galaxies are produced with forward modeling techniques based on their star formation histories. As SNe occur at different rates in galaxies depending on their properties such as stellar mass and star formation rate, we weight the probability of host galaxy association as in \cite{Vincenzi21a}, based on rates from \cite{Wiseman21}. Lastly, the probability of obtaining a spectroscopic redshift is applied as a function of magnitude, as shown in Figure \ref{fig:zeff}.

\begin{figure*}
\gridline{
    \fig{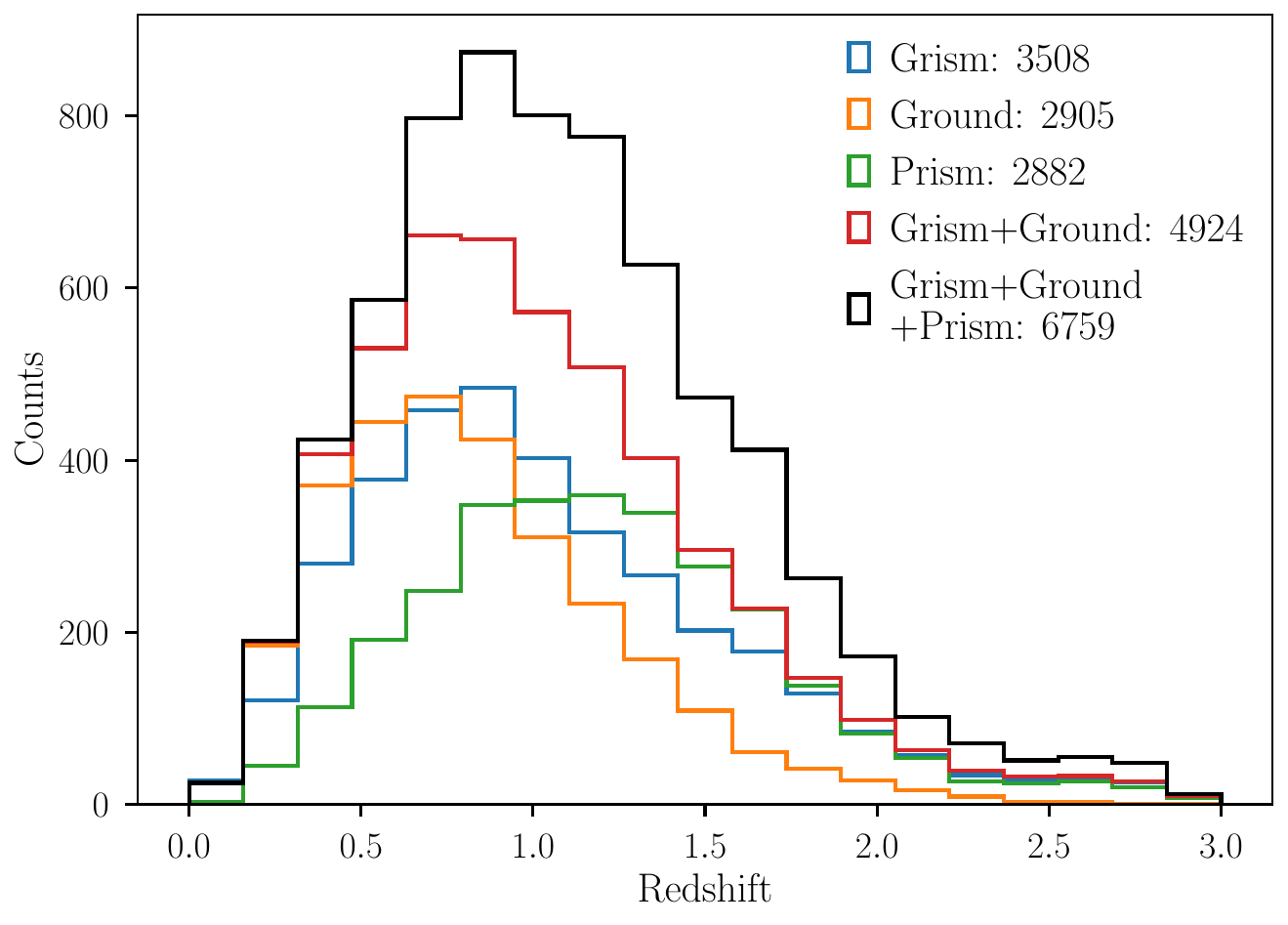}{0.5\textwidth}{}
    \fig{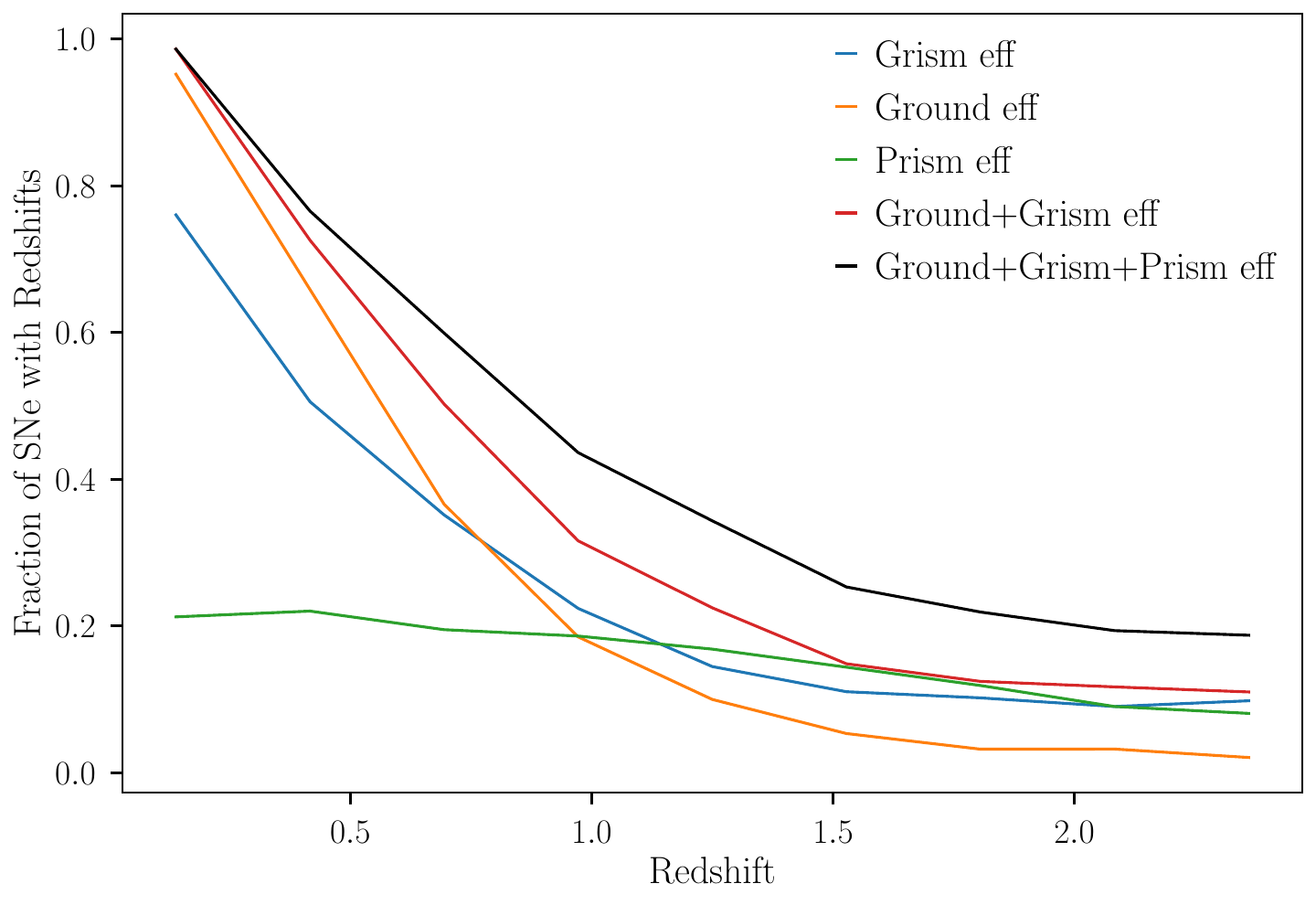}{0.5\textwidth}{}
}
\caption{Left: Post light-curve fit SN redshift distributions for simulations applying a grism efficiency (blue), a ground-based efficiency (orange), a prism efficiency (green), and all three together (black). Right: Overall efficiency for SNe, defined as the number of SNe post light-curve fit for a simulation given a particular redshift efficiency divided by the number of SNe post light-curve fit for a simulation with no redshift efficiency. Note that the prism efficiency is scaled down by 0.25 as described in Section \ref{sec:prismeff}.}
\label{fig:zdists}
\end{figure*}

\subsubsection{Light-curve fitting}
We briefly summarize the framework used to measure distances and construct a cosmological SNIa sample. To fit light-curves, we use the $\chi^2$ minimization fitter implemented in SNANA for the SALT3 model, whereby the model parameters are: $z$ (redshift), $x_{0}$ or $m_{x}$ (overall amplitude, with $m_{x} = -2.5\log_{10}(x_{0})$), $t_{0}$ (time of peak brightness), $x_{1}$ (stretch), and $c$ (color). To measure distances $\mu$, we apply the Tripp formula \citep{Tripp}:

\begin{equation}
    \mu= m_x + \alpha x_1 - \beta c - M -\delta \mu_{\textrm{bias}} + \delta \mu_{\textrm{host}},
    \label{eq:Tripp}
\end{equation}
where $\alpha$ and $\beta$ are standardization nuisance parameters fit for the entire SN sample, $M$ is the absolute magnitude of a SNe with $x_1 = c = 0$, $\delta \mu_{\textrm{bias}}$ is the bias correction computed from simulations (defined in Section \ref{sec:bbc}), and $\delta \mu_{\textrm{host}}$ is a correction applied to account for SN brightness correlation with host-galaxy properties. We use the host-galaxy stellar mass as a host-galaxy property proxy and define the correction as follows:
\begin{equation}
    \delta \mu_{\rm host} = \begin{cases}
+\gamma/2 & \quad \text{if } M_* > 10^{10} M_{\odot}, \\
-\gamma/2  & \quad \text{otherwise,}
\end{cases}
    \label{eq:massstep}
\end{equation}
where $\gamma$ is the size of the `mass step' and $10^{10} M_{\odot}$ is the location of the step \citep{Lampeitl10, Kelly10, Sullivan10}.

The following standard cosmological cuts are then applied:
\begin{itemize}
    \item fitted color $\left| c \right| < 0.3$
    \item fitted stretch $\left| x_1 \right| < 3.0$
    \item fitted stretch uncertainty $\sigma_{x_1} < 1.0$
    \item fitted $t_0$ uncertainty $\sigma_{t_0} < 2.0$ days
\end{itemize}

\subsection{Non-grism redshift efficiencies}
\label{sec:othereff}
As the grism will not be the only source of redshifts for $Roman$ SN cosmology, here we consider two additional sources, the $Roman$ prism and a ground-based spectrograph, to understand how each will contribute to a final SN redshift distribution.

\subsubsection{$Roman$ prism redshift efficiency}
\label{sec:prismeff}

In addition to grism redshifts, a subset of SNe will also be observed in the Southern EDFS field with the $Roman$ prism as part of the main HLTDS. \citet{Joshi22} determine the efficiency of recovering redshifts measured from SN spectra using two-dimensional dispersed images with the \texttt{pyLINEAR} software \citep{Ryan18}. With a requirement of $\sigma_z \leq0.01$, they measure the 50\% redshift F106 magnitude completeness for one hour exposures to be 24.41$\pm$0.06. As it is not yet determined what percentage of survey time will be focused on obtaining live SN spectra vs. host galaxy spectra, for simplicity here we do not differentiate between the two and refer to both as ``spectroscopic redshifts.'' However, given that galaxies are extended sources and have greater diversity in their spectra and narrower spectral features relative to SNeIa, we additionally shift the SN redshift efficiency brighter by 0.5 mag. 

A total of $\sim$20\% of HLTDS time is expected to be allocated to a wide and deep spectroscopy tier. As the area ratio of the Wide Imaging Tier to the Wide Spectroscopy Tier is roughly 25\%, we scale the efficiency for one hour exposures from \citet{Joshi22} by 0.25 to use as our prism efficiency. In other words, the maximum possible recovery rate is scaled down from 1 to 0.25. This is an optimistic approximation, as the recommended in-guide implementation of the HLTDS allocates one hour exposures only for the Deep Tier, with shorter 900 s exposures for the Wide Tier. 

\subsubsection{Ground-based efficiency}
In addition to redshifts from the prism and grism, spectroscopic redshifts will also be obtainable from various ground-based telescopes, such as through partnership with the Subaru Telescope Prime Focus Spectrograph \citep{PFS}, particularly at redshifts $<1$. As the allocation of Subaru PFS $Roman$ time has not yet been finalized, here we use instead the efficiency from the DES-SN5YR analysis \citep{Vincenzi21a, Vincenzi24} as a rough proxy. This efficiency was calculated for the OzDES \citep{Lidman20} spectroscopic follow-up program on the Anglo-Australian Telescope, which aimed for high completeness to a magnitude limit of $m_r=24$. The efficiency was modeled as a function of $r$ band magnitude, $g-r$ color, year of discovery (Years 1-5, as the number of fiber hours in each year differed), and field (as two fields were deep and the rest shallow). We average the efficiency over year of discovery, weighted by the fraction of total observing time, and average only the shallow field efficiencies to obtain an efficiency as a function of $r$-band magnitude and $g-r$ color only. This results in a $90\%$ and $50\%$ efficiency at \groundredninety{} and \groundredfifty{} for $g-r < 1.15$ and \groundblueninety{} and \groundbluefifty{} for $g-r \geq 1.15$ respectively.

\subsection{Final SN efficiencies} \label{sec:final_sn_effs}
Using the host galaxy redshift efficiencies detailed in Section \ref{sec:grismeff} and \ref{sec:othereff} (grism, prism, and ground-based), we generate simulated instances of the HLTDS. 

In Figure \ref{fig:zdists} we show in blue the resulting redshift distribution for SNIa when applying a grism-only efficiency to our simulations. We show the redshift distribution for simulations with a ground-based only efficiency in orange, a prism-only efficiency in green and a combination of the three in black. We estimate $\sim$\zgrismnum\ SN host galaxies will have grism redshifts, $\sim$\zgroundnum\ SN host galaxies will have redshifts from ground-based telescopes, and $\sim$\zprismnum\ SNe or their host galaxies will have prism redshifts. Despite several simplifying assumptions, our prediction for the prism is comparable to the $\sim$2500 SN redshifts estimated in \citet{Rubin22}. While ideally ground-based follow-up resources will also be optimized to observe unique sources, here we allow overlap between each of the redshift sources as follow-up strategies are not yet known. Therefore, the total number of redshifts is smaller than the sum of the grism-only, ground-only, and prism-only efficiency simulations. The resulting number of total $Roman$ SNe with redshifts is $\sim$\specSN, roughly half of the total cosmological sample sizes predicted with previous naive redshift efficiencies. The mean redshift for the grism only, ground-based only, and prism only are \zgrismonly{}, \zgroundonly{}, and \zprismonly{}, respectively. In the right panel of Figure \ref{fig:zdists}, we show the overall ``efficiency'' for SNe, defined as the number of SNe that pass light-curve fit cuts for a simulation with a given redshift efficiency divided by the number of SNe that pass light-curve fit cuts for a simulation with no redshift efficiency. As expected from the galaxy efficiencies, ground-based telescopes are more efficient until $z\sim1$, when the grism becomes more efficient at obtaining redshifts. The prism efficiency is limited by the survey area covered by the Spectroscopy Tiers, but contributes significantly at higher redshifts than the grism, as expected.

\begin{figure}
\centering
\gridline{
    \fig{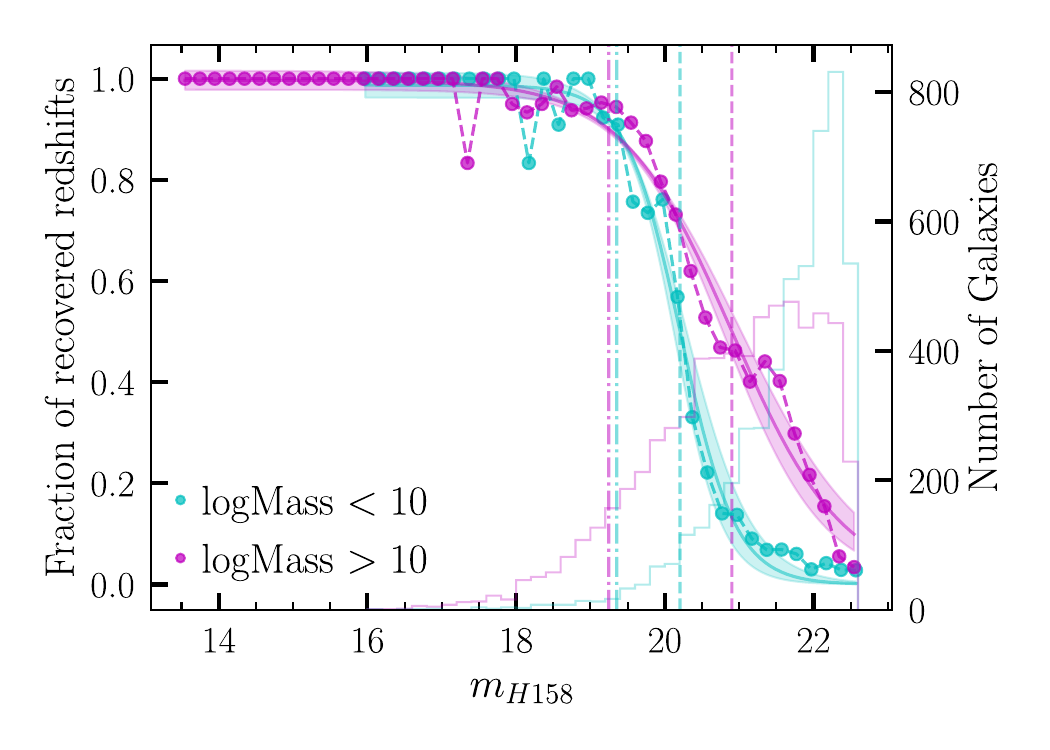}{\columnwidth}{}
}
\vspace{-1.3cm} 
\gridline{
    \fig{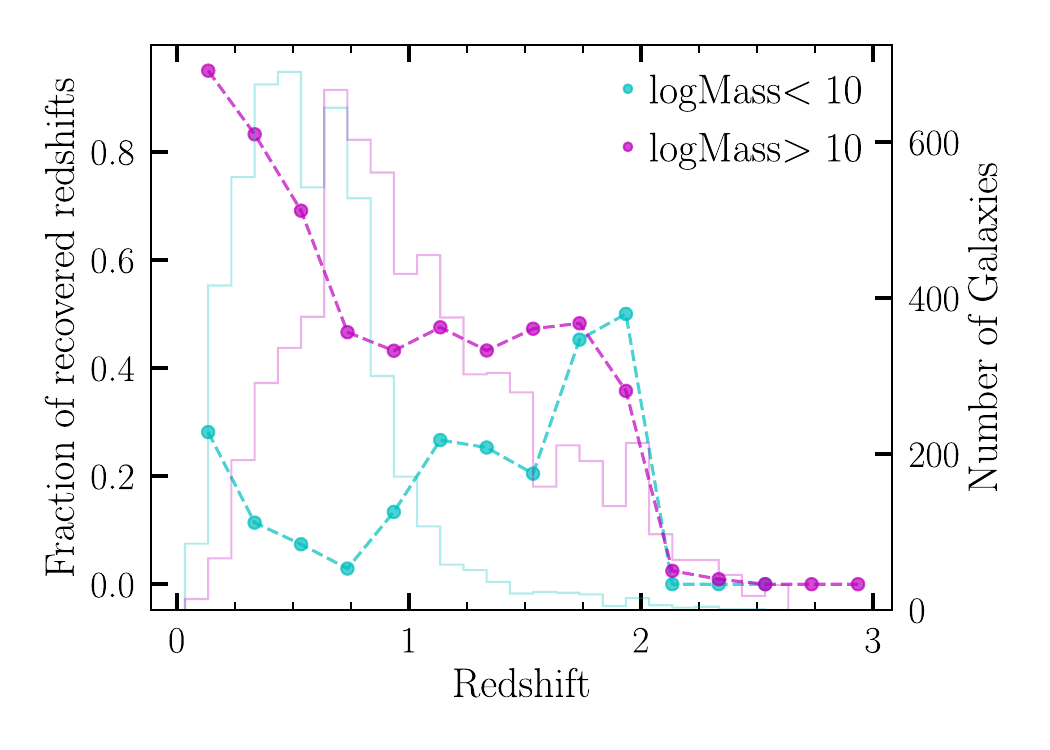}{\columnwidth}{}
}
\vspace{-1cm}
\caption{Top: Same as top of Figure \ref{fig:zeff} but for galaxies with logMass $<10$ in blue and galaxies with logMass $\geq10$ in magenta. Bottom: Same as center of Figure \ref{fig:zeff} but for galaxies with logMass $<10$ in blue and galaxies with logMass $\geq10$ in magenta.}
\label{fig:zeff_mass}
\end{figure}

\begin{figure}
\centering

\gridline{
    \fig{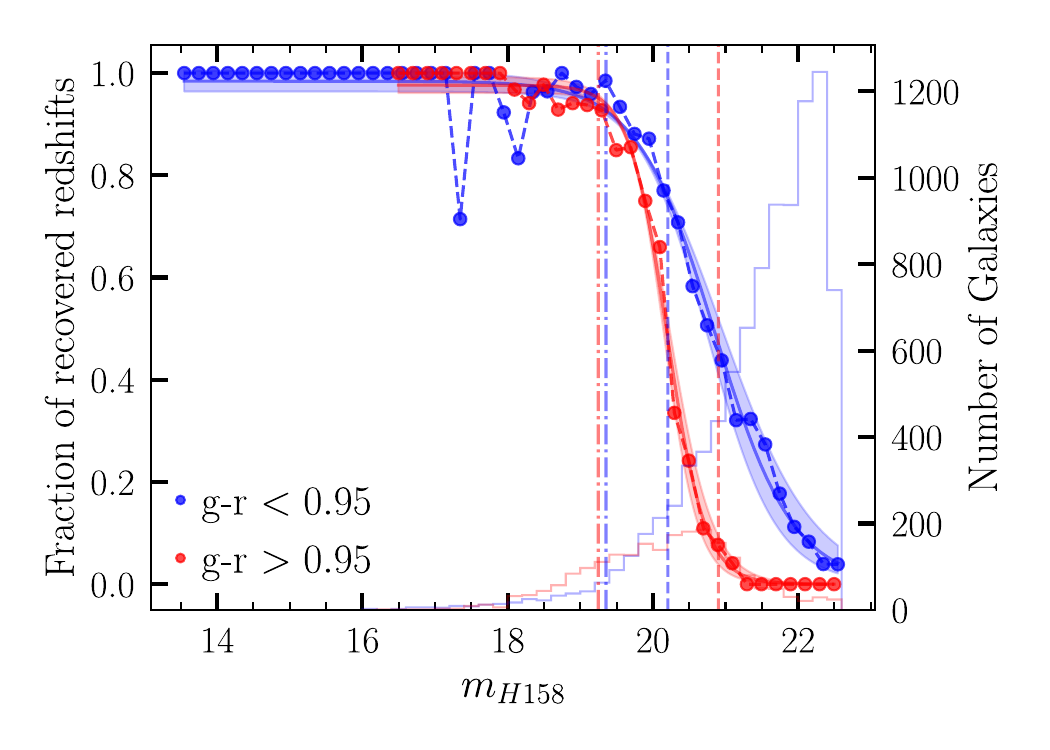}{\columnwidth}{}
}
\vspace{-1.3cm}

\gridline{
    \fig{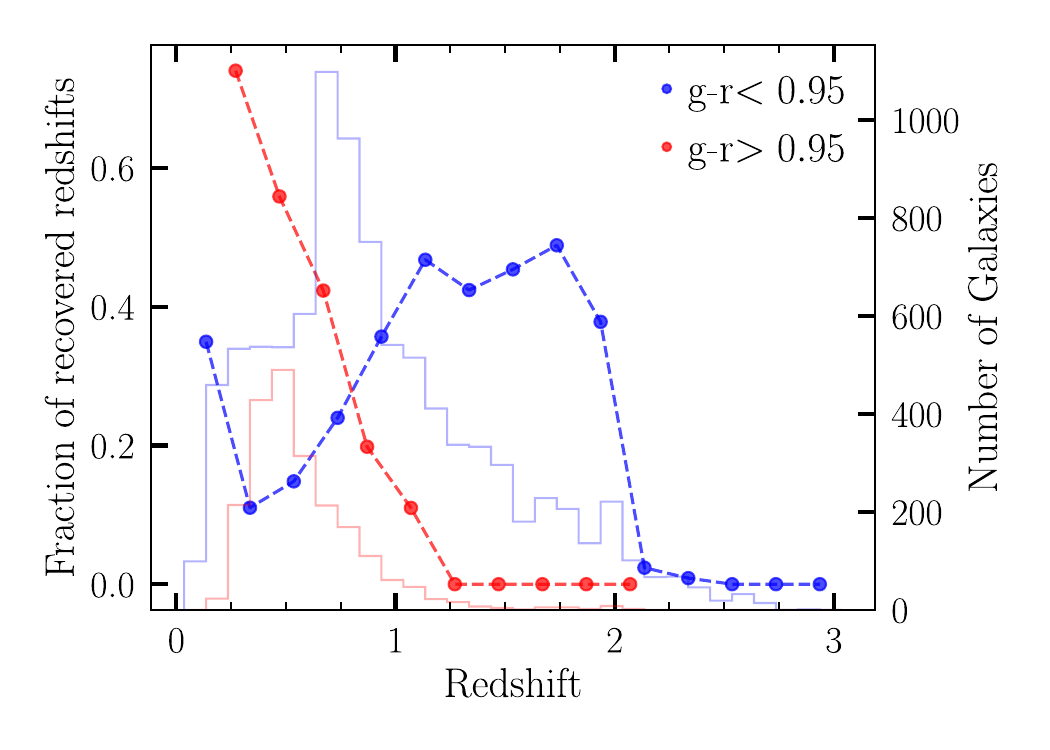}{\columnwidth}{}
}
\vspace{-1cm}
\caption{Top: Same as top of Figure \ref{fig:zeff} but for galaxies with g-r color $<0.95$ in blue and galaxies with g-r color $\geq0.95$ in red. Bottom: Same as center of Figure \ref{fig:zeff_color} but for galaxies with g-r color $<0.95$ in blue and galaxies with g-r color $\geq0.95$ in red.}
\label{fig:zeff_color}
\end{figure}

\section{Systematics}\label{sec:systematics}

An understanding of the spectroscopic redshift efficiency is key in order to accurately model and correct for selection effects in an SNIa cosmology analysis. In Section \ref{sec:bbc} we first detail the cosmological framework used to account for these selection effects and to evaluate their contribution to systematic uncertainties. In Section \ref{sec:eff_modeling} we describe several potential systematics related to the grism redshift efficiency modeling and evaluate their size in Section \ref{sec:eff_systematics}. 

\subsection{Cosmological framework}\label{sec:bbc}

Here we describe the framework used to compute bias corrections, evaluate systematic uncertainties, and constrain cosmological parameters. The BEAMS with bias corrections (BBC; \citealp{BBC}) framework incorporates a cosmology-independent method of fitting for the nuisance parameters $\alpha, \beta, \gamma, \sigma_{\textrm{int}}$ \citep{Marriner11} to produce a binned Hubble diagram. To calculate the $\delta \mu_{\textrm{bias}}$ distance correction term in Eq. \ref{eq:Tripp}, large simulations are generated as described in Section \ref{sec:sn} and used to compute the difference between true and measured distances. For simplicity, we use the BBC 1D formalism, which computes these differences as a function of redshift only. We anchor the Hubble Diagram with a low-$z$ sample consisting of $\sim$700 simulated Foundation \citep{Foley18} supernovae. 

To constrain cosmological parameters, we use the ``{\tt wfit}'' $\chi^{2}$ minimization program as implemented in SNANA with a prior from the Cosmic Microwave Background (CMB) R-shift parameter.

Here we evaluate several potential systematics related to the modeling of the \textit{grism} redshift recovery efficiency, without prism or ground-based redshifts. While a $Roman$ SNIa cosmological analysis will eventually have to account for each source of redshifts, in this work we focus only on the grism. The primary effect of a mismodeled redshift efficiency will be a change in the simulations used to compute bias corrections and correct for selection effects. The variant included in the DES-SN5YR analysis for systematics included a shift of $\pm 0.2$ mag in the efficiency curves. This resulted in a $4\%$ contribution to the total systematic uncertainty budget or shift in $w$ of 0.002.

\subsection{Efficiency modeling}\label{sec:eff_modeling}

Because it is easier to successfully measure redshifts for particular galaxy demographics, e.g. bright, emission line galaxies, and given that SNe are correlated with their host galaxies, it is important to account for such dependencies in the modeling of redshift efficiencies. Here we consider the use of host galaxy stellar mass and host galaxy color as proxies for host galaxy spectral type. In total, we consider three variants for modeling the grism redshift efficiency: i) as a function of host galaxy stellar mass and brightness, ii) as a function of host galaxy color and brightness, and iii) a consistent systematic 0.2 mag shift to the nominal magnitude-only efficiency, as was done in the DES-SN5YR analysis.

To retain sufficient statistics for each population efficiency, we use two mass bins, logMass $<10$ (``low mass'') and logMass $\geq10$ (``high mass''). In Figure \ref{fig:zeff_mass} we show the redshift efficiencies for the low and high mass galaxies as a function of magnitude and redshift. The most notable difference between the two populations is at $z<1$, where the completeness is significantly higher for high mass galaxies compared to low mass. This is consistent with expectations that the grism will primarily target red, passive galaxies at $z<1$. For low mass galaxies we find a 90$\%$ and 50$\%$ efficiency at \lmninety{} and \lmfifty{} respectively, and for high mass galaxies we find a 90$\%$ and 50$\%$ efficiency at \hmninety{} and \hmfifty{} respectively. As expected, the magnitude-only efficiency falls in between the low/high mass galaxy efficiencies consistently at 90 and 50\%. Quantitatively, the efficiency at mag 20 and 21 is \masspercent\ higher for high mass galaxies than low mass galaxies.

Similarly, for galaxy color we split the sample into two populations, ``blue'' galaxies with $g-r$ color $<0.95$, and ``red'' galaxies with $g-r$ color $\geq$0.95, where 0.95 is the value of the 75th percentile for all $g-r$ colors. In Figure \ref{fig:zeff_color} we show the redshift efficiencies for the blue and red galaxies as a function of magnitude and redshift. For blue galaxies we find a 90$\%$ and 50$\%$ efficiency at \blueninety{} and \bluefifty{} respectively, and for red galaxies we find a 90$\%$ and 50$\%$ efficiency at \redninety{} and \redfifty{} respectively. The efficiency at mag 20 and 21 is \colorpercent\ higher for blue galaxies than red galaxies. 

\subsection{Size of systematics}\label{sec:eff_systematics}
As our nominal simulated data, we take the simulations with redshift efficiency as a function of magnitude only. We then compute bias corrections based on simulations with the same magnitude dependent efficiency and measure both $w$ and $w_0$, $w_a$ for 25 data-like statistically independent simulation realizations. We then consider the average shift in the respective parameters when the nominal simulation is bias corrected with each of the three previously described systematic variants. For each variant, we report the shift ($\delta w, \delta w_0, \delta w_a$) in the parameter of interest, along with the robust standard deviation (RSD; 1.48$\times$Median($|\delta|$)) and standard error (RSD/$\sqrt{25}$). The $\delta w$ for each is \masssyst{}, \colorsyst{}, and \shiftsyst{}, with RSD of \masssystRSD{}, \colorsystRSD{}, and \shiftsystRSD{} respectively, compared to the overall expected statistical uncertainty of \wstatuncert{}. 

More relevantly for $Roman$, when fitting instead for $w_0$, $w_a$, we measure a shift in $w_0$ of \womasssyst{}, \wocolorsyst{}, and \woshiftsyst{} respectively, compared to the statistical uncertainty of \wostatuncert{}. For $w_a$, we measure a shift of \wamasssyst{}, \wacolorsyst{}, \washiftsyst{} respectively, compared to the statistical uncertainty of \wastatuncert{}. For $w_0, $ and $w_a$, the mass dependent efficiency variant causes the largest shifts, as is marginally the case for $w$. While for all three variants the $\delta w$ is of comparable size to the 0.002 reported in Table 7 of \citet{Vincenzi24}, the shift in $w_0$ and $w_a$ for the mass dependent efficiency variant is significantly larger than the other two variants. In addition, the relative size of the systematic compared to the overall statistical uncertainty ($\sim$40\%) is much larger than the $\sim$12\% of the other variants. However, this result may be primarily driven by the fact that we consider shifts only for the $Roman$ grism, for which the efficiency differs significantly at $z<1$ as a function of galaxy mass and color. The smaller statistical errors for $Roman$ will require a tighter control of all systematic uncertainties, and future analyses should therefore consider including this mass-dependent efficiency systematic variant, particularly if there are multiple sources of redshifts.

\section{Effects on survey strategy optimization}\label{sec:FOM}

As part of the efforts to define the HLTDS survey strategy \citep{Kessler25, Rubin25}, various survey configurations have been explored to maximize the dark energy figure-of-merit (FoM), which is a proxy for the strength of an evolving dark energy constraint defined by the Dark Energy Task Force \citep{Albrecht06, Wang08}. The FoM is defined as the reciprocal of the $95\%$ confidence level of the constraint in the $w_{0}$-$w_a$ plane. These survey optimization efforts have made different assumptions on whether the final cosmological constraints will be made with photometric redshifts or spectroscopic redshifts. \citet{Kessler25} assume that spec-$z$'s will be available for all SNe at $z<0.3$, with all other redshifts fit from the SN light-curve with a host photo-$z$ prior \citep{Kessler09}, whereas the Fisher forecast of \citet{Rubin25} assumes that all events will have spec-$z$'s. In order to understand the impact of redshift source assumptions on survey optimization results, we compare the relative shifts in FoM for several survey configurations with redshift efficiencies as modeled in Sections \ref{sec:grismeff} and \ref{sec:othereff}. The simulations used for survey optimization differ slightly from the one described in Section \ref{sec:sn} as the model used for SN intrinsic scatter is instead dust-based \citep{BS20, Dust2Dust} with populations measured with the Dust2Dust software \citep{Dust2Dust} and parameters presented in \cite{Vincenzi24}. We follow assumptions and calculations for exposure times as detailed in \citet{Kessler25}. As in the previous section, we include the simulated low-$z$ Foundation sample as an anchor survey. 

\begin{figure*}
\centering
\fig{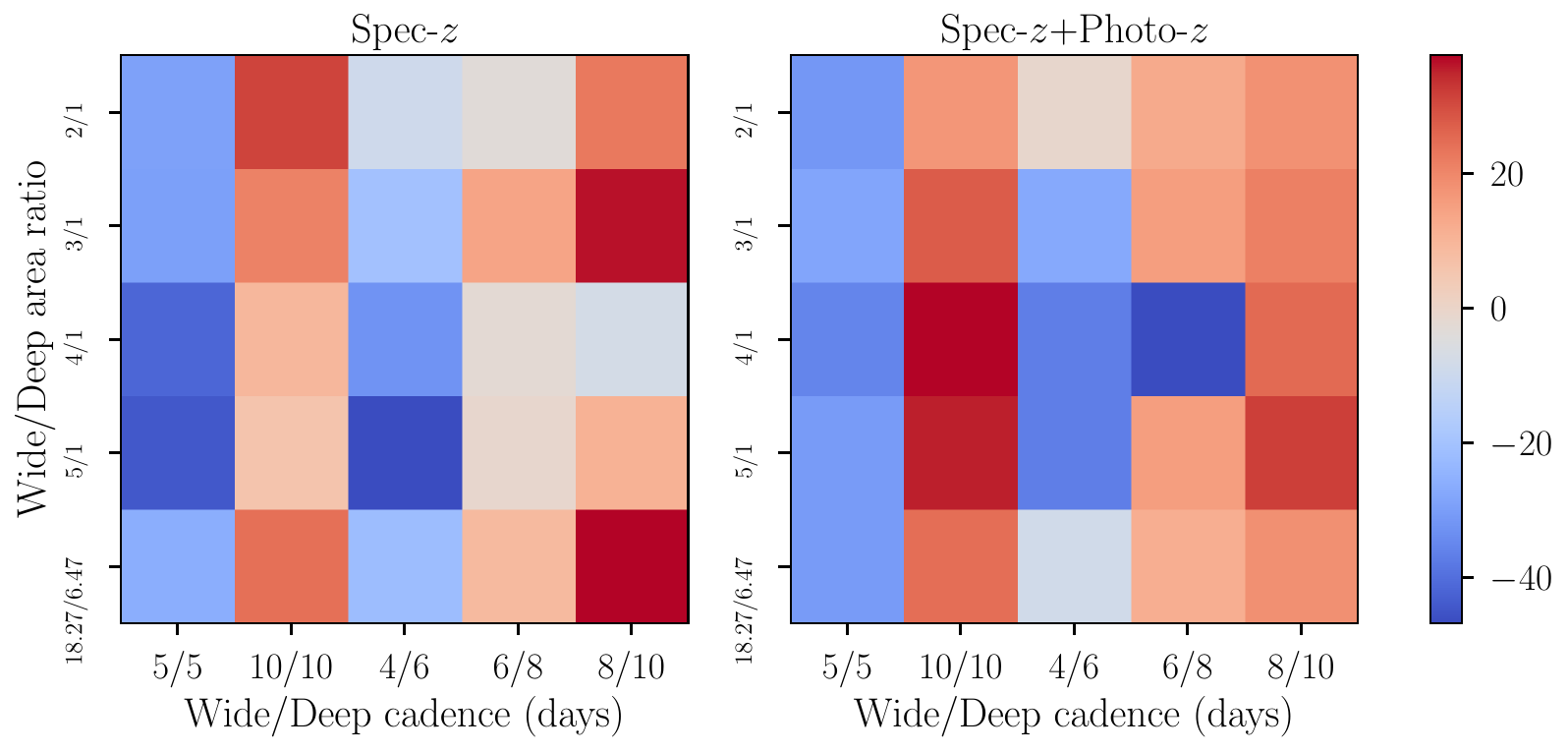}{0.85\textwidth}{}
\caption{Left: Heatmap of FoM values relative to the mean FoM of the grid for varying survey configurations with FoM values calculated with a spec-$z$ requirement, assuming redshifts from ground-based telescopes and $Roman$ grism. Right: Same as left but when allowing for either a spec-$z$ or photo-$z$ fit from the SN with a host galaxy prior.}
\label{fig:fom_gg}
\end{figure*}

\begin{figure*}
\centering
\fig{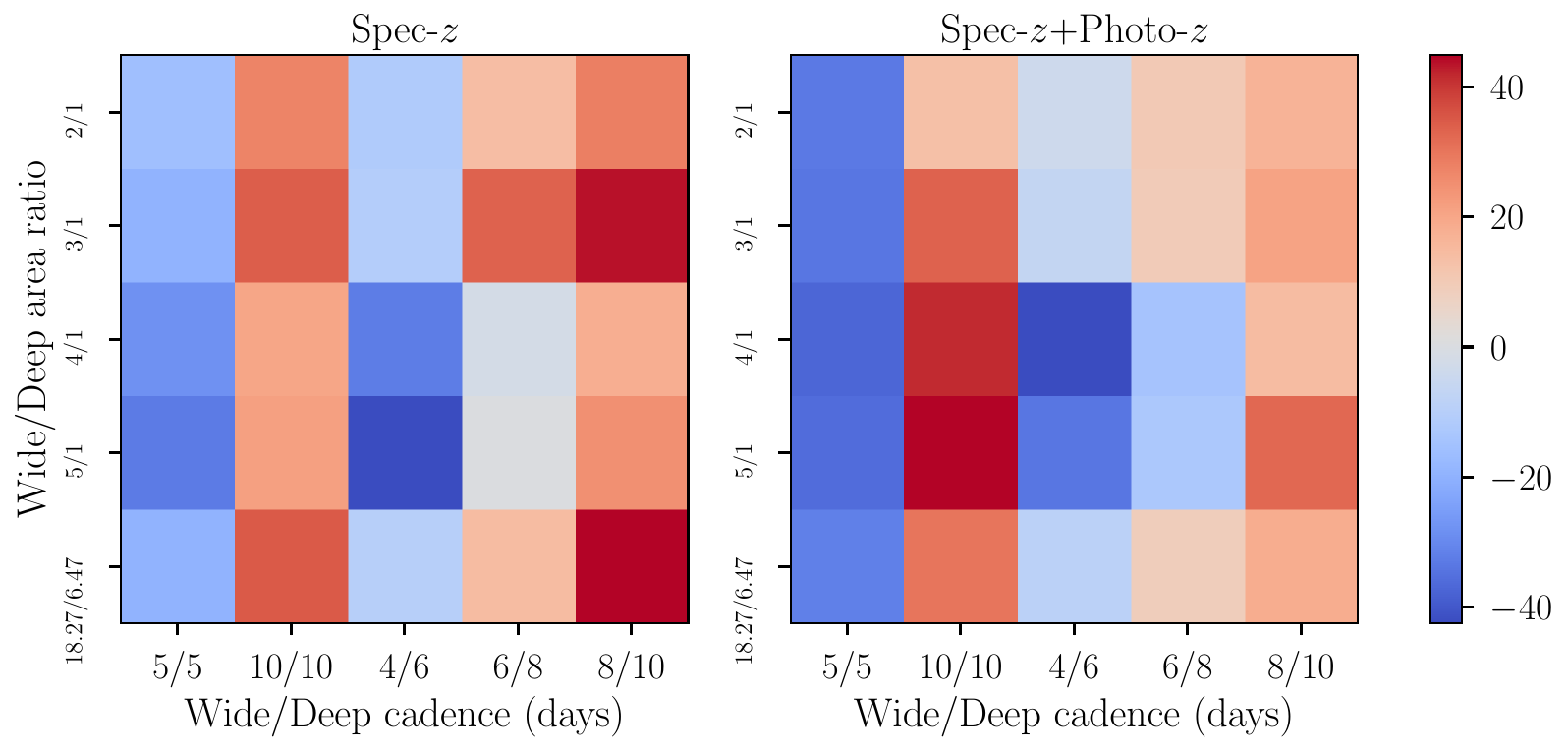}{0.85\textwidth}{}
\caption{Same as Figure \ref{fig:fom_gg} but when including redshifts from ground-based telescopes, $Roman$ grism, and $Roman$ prism.}
\label{fig:fom_ggp}
\end{figure*}

\subsection{Survey Configurations and Figure of Merit}

We consider a simple, non-comprehensive, two-dimensional grid of configurations with five options for i) the relative survey area between the wide and deep surveys and ii) the cadence for the wide and deep surveys, resulting in 25 distinct configurations. We fix the target redshifts to 0.9 and 1.7 for the wide and deep surveys respectively, which are used to determine the exposure time such that the signal-to-noise ratio (SNR) in each band is 10. 

While in this work we have treated host galaxy grism and ground-based telescope redshifts and prism SN redshifts identically as ``spectroscopic redshifts,'' a proper cosmological analysis will need to model and account for each source of redshift separately. We first make a conservative assumption of host galaxy redshifts only, from ground-based and grism efficiencies. In Figure \ref{fig:fom_gg}, we show the FoM values relative to the mean FoM of the grid for simulations generated with each survey configuration. The left grid is calculated with a spectroscopic redshift requirement, and the right grid additionally allows a photo-$z$ fit from the SN light-curve with a host galaxy photo-$z$ prior (SN+host photo-$z$) when spec-$z$'s are not available. As the FoM values are overall higher for the "spec-$z$+photo-$z$" configurations (due to the larger sample size), it is most illustrative to consider how the FoM values change relative to the individual grid mean FoM. These relative FoM values can then be compared directly between the left and right grids. For example, squares which are blue (worse FoM) in the spec-$z$ grid but red (better FoM) in the spec-$z$+photo-$z$ grid correspond to survey configurations which are especially sensitive to the sources of redshifts. In Figure \ref{fig:fom_ggp}, we show the same maps for simulations assuming an efficiency including both host galaxy redshifts from ground-based telescope and the $Roman$ grism, as well as SN redshifts from the $Roman$ prism. 

The 18.27/6.47 Wide/Deep area ratio is the ratio from the final in-guide recommendation by the Core Community Survey Definition Committee. However, we emphasize that none of the configurations considered here are exactly identical to the in-guide recommendation, as details such as an interleaved cadence for the Wide/Deep fields are not included for simplicity. Rather, we highlight that conclusions on optimal survey configurations and their relative FoM values have a non-negligible dependence on the source of redshifts and whether they are spectroscopic or photometric. Therefore, decisions related to survey optimization and optimal host galaxy follow-up must necessarily take into account high level analysis choices, particularly regarding the use of photometric vs. spectroscopic redshifts.

\section{Conclusion}\label{sec:conclusion}

In this work, we have characterized three elements related to obtaining redshifts for the \textit{Roman} HLTDS for SNIa cosmology, focusing primarily on the \textit{Roman} grism. 
\begin{enumerate}
\item We determine the \textit{Roman} grism galaxy redshift recovery rate using simulated 2D spectra on top of the OpenUniverse2024 image simulations and apply it to catalog-level SN simulations of the HLTDS. Further, we consider the overall redshift schema for SNIa cosmology by also generating simulations including an approximate ground-based telescope and $Roman$ prism redshift efficiency.
\item We further consider a grism redshift efficiency modeled as a function of magnitude and host mass as well as magnitude and host color. We evaluate three systematic variants related to this modeling of the redshift efficiency and find shifts in $w$ of comparable magnitude to values reported in the DES-SN5YR analysis. We highlight that while the modeling of spectroscopic redshift efficiency has not been a major systematic for previous cosmological analyses, the increased statistical power of $Roman$ may require a more precise understanding of redshift recovery efficiency for $w$ and $w_0$-$w_a$, as presented in this work.
\item We investigate the degree to which assumptions on both the type and sources of redshifts have an impact on the relative FoM for different HLTDS survey configurations. We emphasize that conclusions regarding optimal survey strategy depend consequentially on whether an analysis assumes usage of photometric redshifts.
\end{enumerate}

Notably, for the grism and ground based efficiencies, here we do not account for the fact that the grism overlaps with the HLTDS only in the south and that Subaru PFS will provide spectra only in the north. Our number estimates thus serve as an optimistic upper bound which illustrate the potential of the grism for the SN cosmology use-case.

While \citet{Kessler25} presents a first end-to-end cosmological analysis of simulated $Roman$ data, including several simple systematics in the covariance matrix formalism, further work remains to provide a complete picture of the systematics that will be most important for $Roman$ SN cosmology. Several simplifying assumptions have been made here which should also be explored in future work, including the treatment of both $Roman$ grism and prism redshifts as exactly equivalent to those obtained from ground-based spectrographs despite their larger redshift scatter. As has been noted in \citet{Kessler25} and \citet{Chen25}, the unbiased use of photometric redshifts relies heavily on accurately simulated redshifts for the bias correction simulation. A more sophisticated analysis will therefore also require modeling of the redshift dispersion specific to each instrument.

\section*{Data Availability}

The \texttt{Pippin} and \texttt{SNANA} configuration files used in this analysis are available upon reasonable request to the authors. 

\begin{acknowledgments}
The authors thank Stefano Casertano for discussion regarding the grism noise model. D.S. is supported by Department of Energy grant DE-SC0010007, the David and Lucile Packard Foundation, the Templeton Foundation, and Sloan Foundation. L.G. acknowledges financial support from AGAUR, CSIC, MCIN and AEI 10.13039/501100011033 under projects PID2023-151307NB-I00, PIE 20215AT016, CEX2020-001058-M, ILINK23001, COOPB2304, and 2021-SGR-01270. Funding for the Roman Supernova Project Infrastructure Team has been provided by NASA under contract to 80NSSC24M0023. This work was completed, in part, with resources provided by the University of Chicago's Research Computing Center.
\end{acknowledgments}

\software{Matplotlib \citep{matplotlib}, 
Numpy \citep{numpy}, 
Pandas \citep{pandas}, 
Pippin \citep{Pippin},
Python, 
SciPy \citep{scipy}, 
\texttt{SNANA} \citep{SNANA},
\texttt{Grizli} \citep{Grizli}}.


\appendix

\section{Relaxed redshift requirements}

In the main body, we present grism redshift efficiencies and approximate expected sample sizes based on strict redshift requirements; however, more relaxed cuts may provide sufficiently precise redshift estimates for applications such as providing a prior for fitting SN+host photo-$z$. Here we present the redshift efficiencies and number of SNe expected as shown in Section \ref{sec:grismeff} and \ref{sec:final_sn_effs} for relaxed redshift requirements including a single line detection and a lower detection threshold.

\begin{figure}
\centering
\fig{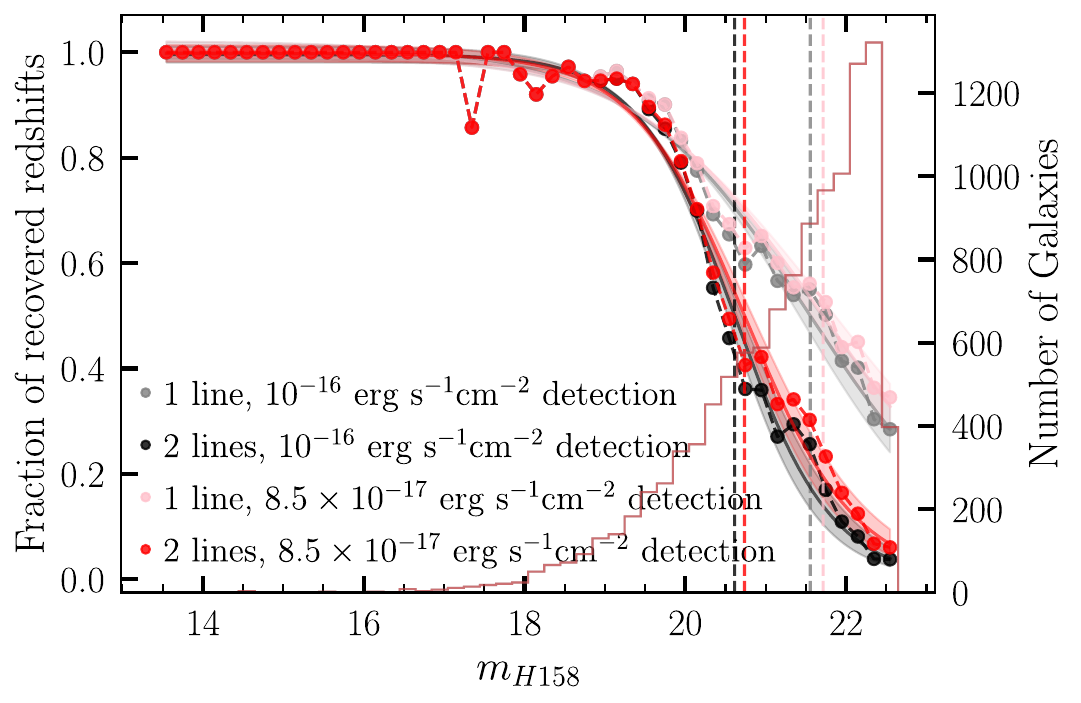}{0.8\columnwidth}{}
\vspace{-0.5cm}
\caption{Redshift efficiency from the grism image simulations as in Figure \ref{fig:zeff} but for relaxed redshift requirements: $10^{-16}$ erg $\textrm{s}^{-1} \textrm{cm}^{-2}$ detection threshold and two or more lines detected (nominal, black); $10^{-16}$ erg $\textrm{s}^{-1} \textrm{cm}^{-2}$ detection threshold and one or more lines detected (grey); $8.5\times 10^{-17}$ erg $\textrm{s}^{-1} \textrm{cm}^{-2}$ detection threshold and two or more lines detected (red); $8.5\times 10^{-17}$ erg $\textrm{s}^{-1} \textrm{cm}^{-2}$ detection threshold and one or more lines detected (pink).}
\label{fig:hosteff_relaxed}
\end{figure}

In Figure \ref{fig:hosteff_relaxed}, we show the corresponding redshift recoveries as a function of magnitude for i) a single line detection compared to two line detection, as well as ii) a detection threshold of $8.5\times 10^{-17}$ erg $\textrm{s}^{-1} \textrm{cm}^{-2}$ (5$\sigma$ depth of the HLWAS), compared to the nominal detection threshold of $10^{-16}$ erg $\textrm{s}^{-1} \textrm{cm}^{-2}$ (6.5$\sigma$ depth of the HLWAS). While each variation results in only a marginal change in the 90\% efficiency, the 50\% efficiency is significantly fainter when the number of lines required to be detected is reduced to one. Compared to the nominal 50\% two line, $10^{-16}$ erg $\textrm{s}^{-1} \textrm{cm}^{-2}$ threshold detection efficiency of 20.61, the 50\% one line, $8.5\times 10^{-17}$ erg $\textrm{s}^{-1} \textrm{cm}^{-2}$ detection is 21.72. 

\begin{figure*}
\centering
\gridline{
    \fig{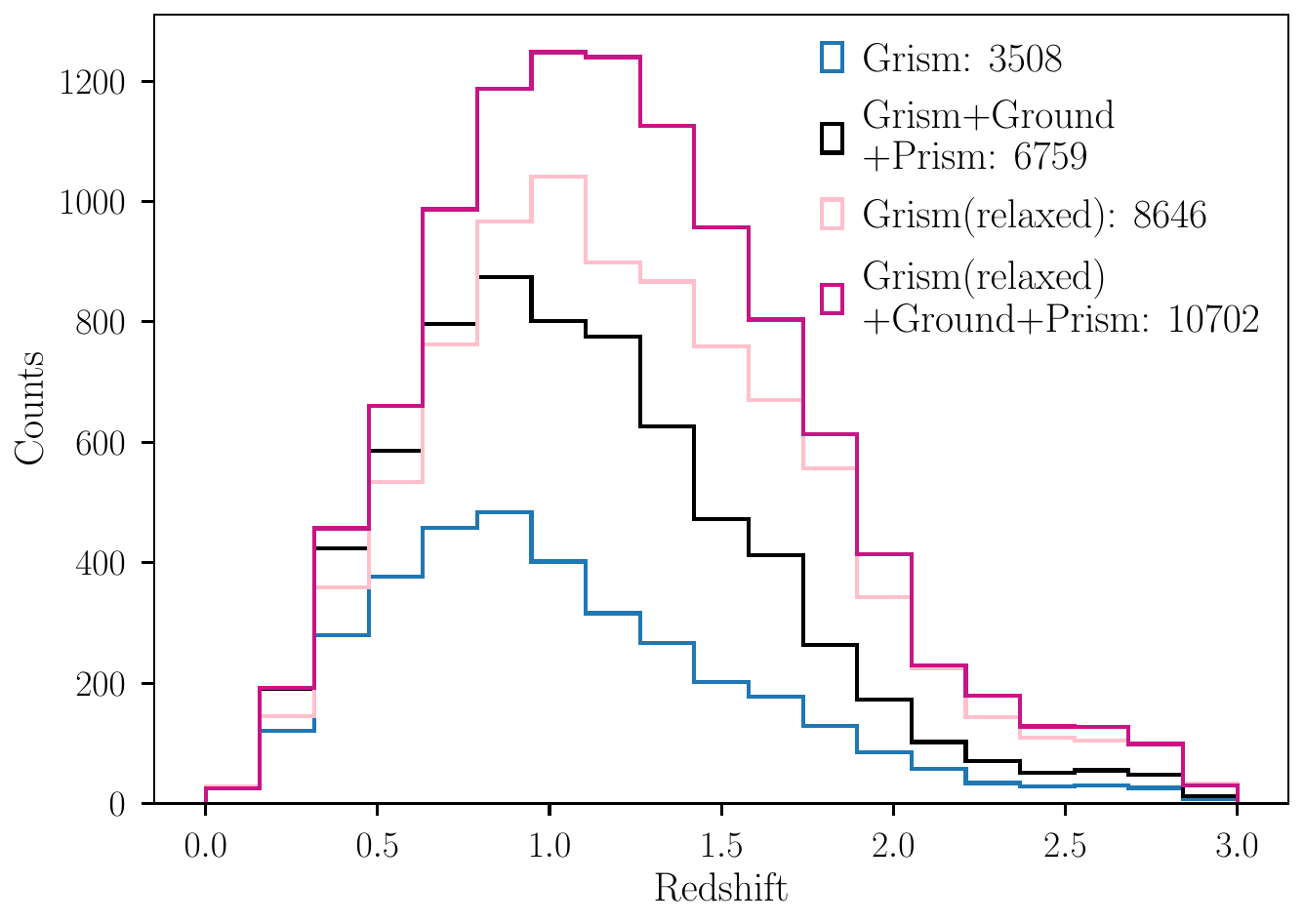}{0.5\textwidth}{}
    \fig{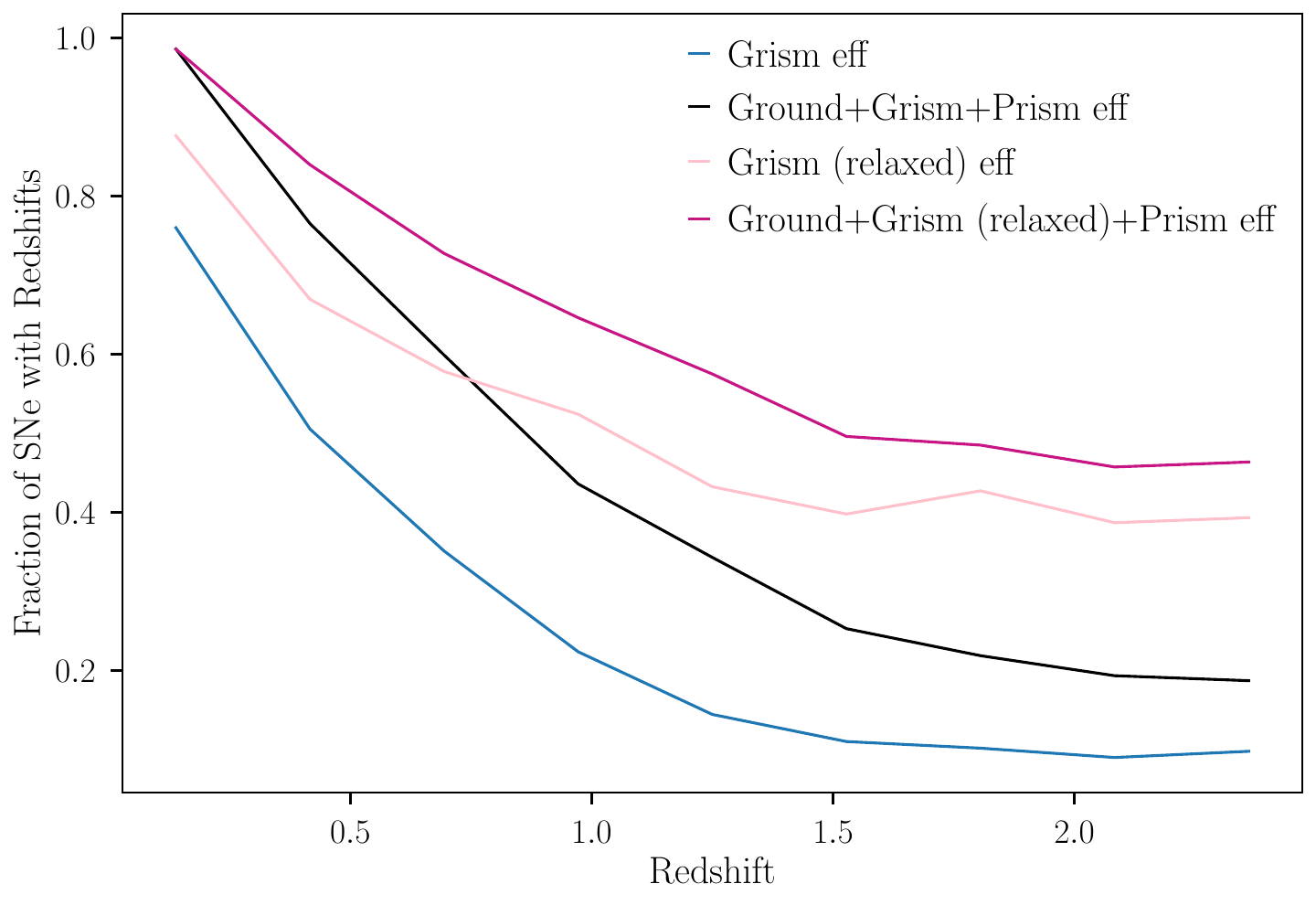}{0.5\textwidth}{}
}
\caption{Same as Figure \ref{fig:zdists}, but for simulations with a relaxed redshift requirement (one line $10^{-16}$ erg $\textrm{s}^{-1} \textrm{cm}^{-2}$ detection threshold).}
\label{fig:zdists_oneline}
\end{figure*}

As the difference between the two detection threshold requirements is minimal, we further consider the expected SN sample sizes only for the nominal one line $10^{-16}$ erg $\textrm{s}^{-1} \textrm{cm}^{-2}$ detection threshold requirement. In Figure \ref{fig:zdists_oneline}, we show the equivalent to Figure \ref{fig:zdists} when relaxing to the one line $10^{-16}$ erg $\textrm{s}^{-1} \textrm{cm}^{-2}$ detection threshold requirement. In total, $\sim11000$ SNe have an associated spectroscopic redshift compared to the nominal $\sim6800$, and $\sim8600$ SNe have a redshift from the grism compared to the nominal $\sim3500$. This increase is primarily at $z<0.5$, as relaxing to a single line detection allows redshifts determined with only an [SIII] emission line to be considered robust. However, these redshifts, as well as most of the redshifts from the grism at $z<0.5$, have much larger typical scatter ($\sim0.01$) than a traditional spec-$z$. While useful for SN population studies or for a combined spec-$z$+photo-$z$ analysis, we note that they are insufficiently precise for a typical spec-$z$ cosmology analysis.

\bibliography{research2}{}
\bibliographystyle{aasjournalv7}



\end{document}